\documentclass{emulateapj}

\usepackage{amssymb}
\usepackage{amsmath}

\def\aj{\rm{AJ}}                    
\def\apj{\rm{ApJ}}                 
\def\apjl{\rm{ApJ}}                
\def\apjs{\rm{ApJS}}                       
\def\mnras{\rm{MNRAS}}   

\usepackage{epsfig}
\shorttitle{A Uniform History for Galaxy Evolution}
\shortauthors{Steinhardt \& Speagle}

\begin{document}


\title{A Uniform History for Galaxy Evolution}


\author{Charles. L. Steinhardt \altaffilmark{1,2,3}, Josh S. Speagle\altaffilmark{3,4}}

\altaffiltext{1}{California Institute of Technology, MC 105-24, 1200 East California Blvd., Pasadena, CA 91125, USA}
\altaffiltext{2}{Infrared Processing and Analysis Center, California Institute of Technology, MC 100-22, 770 South Wilson Ave., Pasadena, CA 91125, USA}
\altaffiltext{3}{Kavli IPMU, University of Tokyo, Kashiwanoha 5-1-5, Kashiwa-shi, Chiba 277-8583, Japan}
\altaffiltext{4}{Harvard University Department of Astronomy, 60 Garden St., MS 46, Cambridge, MA 02138, USA}


\begin{abstract}
Recent observations indicate a remarkable similarity in the properties of evolving galaxies at fixed mass and redshift, prompting us to consider the possibility that most galaxies may evolve with a common history encompassing star formation, quasar accretion, and eventual quiescence.  We quantify this by defining a “synchronization timescale” for galaxies as a function of mass and redshift that characterizes the extent to which different galaxies of a common mass are evolving in the same manner at various cosmic epochs.  We measure this synchronization timescale using 9 different star-forming galaxy observations from the literature and SDSS quasar observations spanning $0 < z \lesssim 6$.  Surprisingly, this synchronization timescale is a constant, approximately 1.5 Gyr for all combinations of mass and time.  We also find that the ratio between the stellar mass of galaxies turning off star formation and black hole mass of turnoff quasars is approximately 30:1, much lower than the 500:1 for quiescent galaxies at low redshift.  As a result, we propose a model in which the star-forming ``main sequence'', analogous quasar behavior, and other observations form a galactic evolution ``main sequence'', in which star formation occurs earliest, followed by supermassive black hole accretion, and feedback between the two are dominated by deterministic rather than stochastic processes.
\end{abstract}

\keywords{galaxies: evolution}


\section{Introduction}


Recent advances have produced a consensus observational result that developing galaxies of a common mass are remarkably similar to each other at any fixed redshift over a broad redshift range.  For example, observations of star-forming galaxies over the past decade have found that there is a very tight relationship (the star-forming ``main sequence''; SFMS) between stellar mass ($M_*$) and star formation rate (SFR, or $\psi$) at all fixed redshifts $0 < z < 6$ \citep{Brinchmann2004,Salim2007,Noeske2007,Elbaz2007,Daddi2007,Ilbert2010,Rodighiero2011,Whitaker2012,Wuyts2011,Karim2011,Zahid2012,Santini2009,Lee2012}.  These studies use a variety of different selection criteria and star formation rate indicators, yet there is strong agremeent between different studies at the same redshift \citep{Speagle2014}.  A corresponding main-sequence-like behavior is found, also at fixed redshift, between the black hole mass ($M_\textrm{BH}$) and bolometric luminosity ($L$) of quasars (cf. \citet{Steinhardt2010a,Steinhardt2011}).  

In order for this main sequence to remain tight at all redshifts, galaxies that have similar stellar mass and SFR at a given redshift must continue to use gas at similar rates, because otherwise the main sequence would see increased scatter towards lower redshift.  As a result, we are motivated to consider that galaxies may actually have quite similar histories, rather than grow stochastically through random bursts of increased activity.  An obvious next step, then, is to better understand that history, with the ultimate goal of developing a {\em Galactic Evolution Main Sequence} analogous to the Hubble Sequence for its morphological evolution.

We first seek to quantify the extent to which these main sequences indicate that different galaxies are being assembled at the same time.  Such a measure has only recently become possible, because it requires comparing views of developing galaxies across a wide range of redshifts.  \citet{Speagle2014} use 64 redshift ranges over 25 different studies to produce a consistent description of the time evolution of the star-forming galaxy population, while the broad redshift coverage of the Sloan Digital Sky Survey makes this comparison more straightforward for quasars.  

In \S~\ref{sec:synchron}, we show how these relations can be used to define a {\em synchronization timescale} ($\tau_s$) for galactic evolution, a new measure of the extent to which different galaxies are being assembled at the same time, one that can be both found observationally for a variety of processes and computed theoretically for different models.  
Because $\tau_s$ is truly a physical timescale associated with each process, it can be easily compared across different types of observations, different physical processes, and different redshifts.  It has previously been difficult to directly compare these relations, because it is uncertain how to relate, e.g., the approximately 0.25 dex (1$\sigma$) in star formation rate at $z \sim 0.1$ for star-forming galaxies \citep{Salim2007} and the approximately 0.8 dex width of the quasar distribution \citep{Steinhardt2010b} at similar redshifts.

We then calculate the synchronization timescale for star-forming galaxies in \S~\ref{sec:sf}, finding that $\tau_s$ is consistent with a constant value at all observed redshifts, $0 < z \lesssim 6$.  \S~\ref{sec:qso} contains the same calculation for quasars, finding a constant $\tau_s$ similar to that for star formation.  Thus, we find that there is indeed strong, quantitative evidence that typical galaxies follow a common evolutionary track, encompassing rapid star formation on the star-forming main sequence, rapid growth of the supermassive black hole as a quasar, and eventual quiescence.  

The natural next step, then, is to understand what that track entails, and in particular whether star formation and quasar activity happen concurrently or sequentially.  In \S~\ref{sec:feedback}, we use galaxies turning off these respective main sequences to try and answer that question, finding that at fixed redshift, there is approximately a 30:1 ratio between the stellar mass of galaxies turning off the star-forming main sequence and the black hole masses of turnoff quasars.  This is substantially smaller than the 500:1 ratio observed in quiescent, $z \sim 0$ galaxies \citep{Haering2004,Magorrian1998}, and provides a new, sharp puzzle.  Finally, we propose one possible Galactic Evolution Main Sequence that may be capable of solving this problem in \S~\ref{sec:discussion}.  Our model is likely just one of many possible sequences for galactic evolution, and we propose a series of observational tests that would be capable of falsifying our model and distinguishing it from other possibilities.

All results listed are derived assuming a \citet{kroupa2001} IMF (integrated from 0.1-100 $M_\odot$) and a $(h,\Omega_m,\Omega_\Lambda) = (0.7,0.3,0.7)$ cosmology.

\section{Synchronization Timescales for Star Formation}

\subsection{Synchronization Timescales}
\label{sec:synchron}

Although this is intended as a general method for considering the views of galactic evolution provided by different processes within a galaxy, it is most instructive to begin by defining $\tau_s$ for one, specific process.  Thus, for simplicity, in this section we will describe everything in terms of studies of star-forming galaxies, leaving other calculations to later sections.

The main sequence of star-forming galaxies is described a tight correlation between $M_*$ and SFR at fixed redshift (i.e. at fixed time $t$).  Because there is a strong redshift dependence, the main sequence should be properly considered as a narrow locus in three dimensions ($M_*$, SFR, $t$).  We wish to consider $\tau_s$ as the (1$\sigma$) scatter about the main sequence in a time-like direction.  Note that $\tau_s$ is therefore also a measure of whether star formation is a good clock: it indicates how well one can recover the redshift of a star-forming galaxy given only $M_*$ and SFR (see also \S~5.3 of \citet{Speagle2014}).  If all galaxies start from a similar state at the same time, $\tau_s$ would be a timescale associated with star formation, but otherwise may be an indicator of a combination of processes in galactic assembly and later evolution.

If every study could be complete for star-forming galaxies over a wide redshift range, calculating $\tau_s$ would be precisely that straightforward.  However, studies typically only observe star forming galaxies over a narrow redshift range, often spanning a shorter time than $\tau_s$.  Thus, this calculation requires combining many studies at different redshifts in order to estimate $\tau_s$.

On average, the specific (log) star formation rate ($\equiv \Psi/M_*$) of galaxies decreases towards lower redshift, meaning that at fixed stellar mass $M_*$, the average log SFR ($\overline{\Psi}(M_*,t)$) decreases monotonically as a function of time.  We use a best-fit approximation of that decline drawn from over two dozen individual studies \citep{Speagle2014} to estimate $\overline{\Psi}(M_*,t)$.  An individual measurement of the scatter about the main sequence in SFR ($\sigma_{SF}$), then, can be turned into the scatter in a time-like direction instead as
\begin{equation}
\tau_{s,\textrm{SF}}(M_*,t) = \frac{\sigma(M_*,t)}{\dot{\overline{\Psi}}(M_*,t)},
\label{eq:tausfr}
\end{equation}
where the derivative $\dot{\overline{\Psi}}$ is with respect to time.  We perform this calculation using recent studies of star forming galaxies in \S~\ref{sec:sf}.  A corresponding calculation can be defined for other physical processes that have similar main sequence-like behavior, as we do for quasars in \S~\ref{sec:qso}.

We note that this new metric introduces uncertainties that are important for this calculation that did not exist for other measures.  We choose to consider the scatter in $\log$ SFR rather than SFR because the distribution in SFR at fixed mass is approximately normal.  In finding a best-fit $\overline{\Psi}(M_*,t)$, there is insufficient data to do a full, three-dimensional analysis.  \citet{Speagle2014} chose to perform a series of fits at fixed $M_*$, which is a sensible choice because mass has been found to be the most important factor in determining the assembly rate of stars \citep{Peng2010} and supermassive black holes \citep{Steinhardt2010b} (see also \citet{GarnBest2010}).  For star formation, different choices here produce similar results, but this may not be the same for other processes.

A more substantial problem is that in order to understand whether galaxies evolve along convergent or divergent tracks, it is necessary to identify sets of galaxies at different redshifts as being part of the same population.  Otherwise, it will be unclear whether $\tau_s$ is increasing or decreasing for a given population over time.  We might wish to pick, e.g., all galaxies with the same halo mass, but for most galaxies that is difficult to measure directly, and thus must be inferred from other properties.

As described in the following sections, we find that $\tau_s$ is consistent with being the same, constant value for all combinations of (stellar) mass and redshift, which means that regardless of how galaxies are matched across redshifts, $\tau_s$ will be constant.  However, as discussed in \S~\ref{sec:discussion}, this also makes it far more difficult to turn these results into a model for galactic evolution.

\subsection{Studies of the Star Forming Main Sequence}

The relationship between SFR and $M_*$ has been reported in 64 redshift ranges over 25 different studies \citep{Speagle2014}.  Although different studies have employed different selection and different techniques for estimating $M_*$ and star formation rates, \citet{Speagle2014} transforms each study to a common set of calibrations, finding good agreement between different studies ($\lesssim 0.2$ dex scatter between SFR at fixed mass between publications). Every measurement shows a tight correlation between star formation rate and stellar mass, suggesting that it is reasonable to take the set of galaxies at fixed stellar mass $M_*$ as an ensemble.  For each of the measurements, \citet{Speagle2014} calculate the average log SFR $\overline{\Psi}(M_*)$ for all $M_*$ with sufficient statistics and completeness.

Many of these studies report the average properties of a stacked sample of galaxies.  Only 18 of these studies comprising 38 total measurements use individual galaxies and thus are suitable for determining synchronization timescales.  Of these 38, 9 measurements (9 studies) use selection criteria shown in \citet{Speagle2014} to give a restricted view of the SFMS\footnote{Note that these criteria are \textit{not} equivalent to the ``UV''-selection grouping from \citet{Speagle2014} -- see Table \ref{table:sfr} for more details.} (and thus systematically underestimate $\tau_s$), while 11 measurements (2 studies) use selection criteria more appropriate to studying ``average'' galaxy evolution rather than the SFMS (and thus systematically overestimate $\tau_s$), leaving us with 17 measurements that accurately measure the SFMS (see Table~3 and Fig.~1 in \citet{Speagle2014}).  

Using the same (often overlapping) cuts as described in \citet{Speagle2014} -- excluding data (individual observations) where the selection criteria gives a restricted view of the SFMS and would underestimate the scatter (3), the selection criteria gives too broad a view of the SFMS and would overestimate the scatter (11), the observation(s) are based upon $< 250$ galaxies (2), the observations include a strongly biased source population (1), the observations include stacked data in best fit (1), the center of the reported redshift distribution lies between $t < 2$ or $t > 11.5$ Gyr (9), and/or the best reported SFMS includes a sigma-clipping procedure (4) -- we are left with a highest quality sample of 9 measurements (Table \ref{table:sfr}).  These studies report the standard deviation of $\Psi$, but typically do not include a full catalog of individual objects.  We use this scatter to calculate $\tau_s(M_*,t)$ as defined in \S~\ref{sec:synchron}.

\subsection{Synchronization Timescales for Star Formation}
\label{sec:sf}

We now use these high-quality studies to calculate the synchronization timescale for star formation, $\tau_s$, as defined in Eq. \ref{eq:tausfr}.  The behavior at fixed mass is considered first because of its utility as an intermediate step in calculating $\tau_s$.

For $\dot{\overline{\Psi}}$, we use the result from \citet{Speagle2014} that the time evolution of $\overline{\Psi}$ at fixed stellar mass is well-fit by the log-linear
\begin{equation}
\overline{\Psi}(\textrm{fixed }M_*,t) = \alpha - \beta t,
\end{equation}
with a different $\alpha$ but similar slope $\beta \sim -0.16$ dex per Gyr at every mass.  It should be noted that the evolution of $\Psi$ {\em at fixed mass} does not describe the SFR evolution of any individual galaxy.  By definition, the stellar mass of a galaxy increases during star formation, so that the average galaxy does evolve along the two-dimensional surface $\overline{\Psi}(M_*,t)$ but with changing $M_*$.  

\begin{deluxetable*}{l c c c c c c c c}
\tablewidth{0pt}
\setlength{\tabcolsep}{2pt}
\tabletypesize{\footnotesize}
\tablecaption{Synchronization Timescales $\tau_o$ (measured $sigma$), $\tau_d$ ($z$ range corr.), and $\tau_t$ ($z$ and SFR err. corr.) \label{tab:synchro}}
\tablehead{
\colhead{Paper} &
\colhead{$z$} & 
\colhead{$t$ (Gyr)} & 
\colhead{$\log M_*$ ($M_\odot$)} & 
\colhead{$\tau_{s,o}$ (Gyr)} & 
\colhead{$\tau_{s,d}$ (Gyr)} & 
\colhead{$\tau_{s,t}$ (Gyr)} &
\colhead{$\sigma\tau_{s,t}$ (Gyr)} &
\colhead{Exclusion Criteria}
} 
\startdata
Whitaker+12\nocite{Whitaker2012} & 0.25 & 10.52 & 10.0 & 2.373 & 2.072 & 1.532 & 0.156 & None \\
Noeske+07\nocite{Noeske2007} & 0.45 & 8.789 & 10.5 & 2.189 & 1.97 & 1.522 & 0.138 & None  \\
Whitaker+12 & 0.75 & 6.89 & 10.2 & 2.241 & 2.113 & 1.651 & 0.148 & None \\
Zahid+12\nocite{Zahid2012} & 0.785 & 6.71 & 10.0 & 1.884 & 1.881 & 1.261 & 0.151 & None \\
Salmi+12\nocite{salmi+12} & 0.9 & 6.166 & 10.5 & 2.001 & 1.752 & 1.227 & 0.133 & None \\
Elbaz+07\nocite{Elbaz2007} & 1.0 & 5.747 & 9.9 & 2.083 & 2.026 & 1.475 & 0.158 & None \\
Whitaker+12 & 1.25 & 4.875 & 10.5 & 2.126 & 2.084 & 1.667 & 0.14 & None \\
Whitaker+12 & 1.75 & 3.658 & 10.8 & 1.985 & 1.97 & 1.586 & 0.134 & None \\
Whitaker+12 & 2.25 & 2.866 & 11.0 & 1.95 & 1.943 & 1.568 & 0.137 & None \\
--- \\
Elbaz+11\nocite{Elbaz2011} & 0.05 & 12.788 & 10.3 & 1.687 & 1.667 & 1.047 & 0.262 & 4,6 \\
Elbaz+07 & 0.06 & 12.659 & 10.2 & 1.645 & 1.631 & 0.987 & 0.247 & 6 \\
Zahid+12 & 0.07 & 12.532 & 9.5 & 2.034 & 2.019 & 1.195 & 0.341 & 6 \\
Coil+14 (in prep.)\nocite{Moustakas2013} & 0.1 & 12.161 & 10.2 & 4.285 & 4.256 & 4.047 & 0.31 & 2,6 \\
Salim+07\nocite{Salim2007} & 0.11 & 12.04 & 10.1 & 2.001 & 1.926 & 1.389 & 0.273 & 6 \\
Coil+14 (in prep.) & 0.25 & 10.52 & 10.0 & 3.699 & 3.687 & 3.412 & 0.319 & 2 \\
Coil+14 (in prep.) & 0.35 & 9.598 & 10.1 & 3.401 & 3.392 & 3.119 & 0.297 & 2 \\
Sobral+14\nocite{sobral+14} & 0.4 & 9.18 & 9.5 & 3.987 & 3.987 & 3.64 & 0.448 & 2 \\
Coil+14 (in prep.) & 0.45 & 8.789 & 10.3 & 3.373 & 3.366 & 3.106 & 0.284 & 2 \\
Santini+09\nocite{Santini2009} & 0.45 & 8.789 & 9.1 & 2.448 & 2.223 & 0.642 & 0.445 & 7 \\
Coil+14 (in prep.) & 0.575 & 7.912 & 10.4 & 3.226 & 3.215 & 2.955 & 0.277 & 2 \\
Coil+14 (in prep.) & 0.725 & 7.023 & 10.6 & 2.906 & 2.898 & 2.621 & 0.267 & 2 \\
Santini+09 & 0.8 & 6.635 & 9.4 & 1.664 & 1.501 & 1.501 & 0.364 & 7 \\
Sobral+14 & 0.845 & 6.417 & 10.4 & 1.581 & 1.581 & 0.948 & 0.256 & 2 \\
Coil+14 & 0.9 & 6.166 & 10.8 & 2.978 & 2.97 & 2.73 & 0.26 & 2 \\
Santini+09 & 1.25 & 4.875 & 9.6 & 2.357 & 2.3 & 1.726 & 0.328 & 7 \\
Sobral+14 & 1.466 & 4.28 & 10.4 & 1.455 & 1.455 & 0.718 & 0.254 & 2 \\
Kashino+13\nocite{kashino+13} & 1.55 & 4.079 & 10.5 & 1.376 & 1.363 & 0.541 & 0.251 & 1,3 \\
Daddi+07\nocite{Daddi2007} & 1.95 & 3.303 & 10.1 & 1.534 & 1.434 & 0.525 & 0.268 & 1 \\
Zahid+12 & 1.985 & 3.246 & 10.0 & 1.675 & 1.566 & 0.71 & 0.281 & 3 \\
Rodighiero+11\nocite{rodighiero+11} & 2.0 & 3.223 & 10.5 & 1.501 & 1.434 & 0.701 & 0.252 & 1 \\
Santini+09 & 2.0 & 3.223 & 9.9 & 2.569 & 2.521 & 2.104 & 0.298 & 7 \\
Reddy+12\nocite{reddy+12b} & 2.05 & 3.146 & 10.0 & 2.582 & 2.528 & 2.107 & 0.295 & 5 \\
Sobral+14 & 2.23 & 2.892 & 10.6 & 1.484 & 1.484 & 0.82 & 0.249 & 2 \\
Magdis+10\nocite{magdis+10} & 3.0 & 2.109 & 10.9 & 1.216 & 1.214 & 0.362 & 0.232 & 3 \\
Lee+12\nocite{Lee2012} & 3.9 & 1.559 & 9.3 & 2.561 & 2.555 & 1.901 & 0.399 & 6 \\
Shim+11\nocite{Shim2011} & 4.4 & 1.349 & 9.7 & 1.688 & 1.682 & 0.922 & 0.287 & 6 \\
Lee+12 & 5.1 & 1.124 & 9.4 & 2.496 & 2.493 & 1.856 & 0.38 & 6 \\
Steinhardt+14 (subm.) & 5.0 & 1.152 & 10.2 & 1.582 & 1.573 & 0.857 & 0.266 & 6
\enddata
\tablecomments{Col. 1: Papers from which SFMS data are drawn (see \citet{Speagle2014}). Col. 2: Center of redshift range reported. Col. 3: Age of the Universe in Gyr at the center redshift reported. Col. 4: The center of the $\log M_*$ distribution, taken from the ranges reported in \citet{Speagle2014} and used to find the appropriate slope when calculating $\tau_s$. Col. 5: Synchronization timescales calculated using reported observed scatters ($\tau_{s,o}$). Col. 6: Synchronization timescales calculated using scatters that have been deconvolved with the width of their respective time bins ($\tau_{s,d}$) and taken from \citet{Speagle2014}. Col. 7: Synchronization timescales calculated using ``true'' scatters ($\tau_{s,t}$) taken from \citet{Speagle2014}. Col. 8: Calculated 1$\sigma$ errors in $\tau_s$. as well as intrinsic calculated using reported observed scatters ($\tau_{s,o}$). Col. 9: Exclusion criteria, which are as follows: 1 -- selection criteria gives a restricted view of the SFMS and would underestimate the scatter (3), 2 -- selection criteria gives too broad a view of the SFMS and would overestimate the scatter (11), 3 -- observation(s) based upon $< 250$ galaxies (2), 4 -- observations which include a strongly biased source population (1), 5 -- includes stacked data in best fit (1), 6 -- data at $t < 2$ or $t > 11.5$ Gyr (8), 7 -- fit includes sigma-clipping procedure (4).}
\label{table:sfr}
\end{deluxetable*}

Each of the studies in Table \ref{table:sfr} finds that $\sigma(M_*)$ is nearly mass-independent, instead reporting one value of $\sigma$ for all masses. However, the best-fit $\dot{\overline{\Psi}}(M_*,t)$ is actually mass-\textit{dependent} \citep{Speagle2014}. In order to calculate the approximate mass-averaged $\tau_{s,\textrm{SF}}(t)$ for each study, we divide the reported $\sigma$'s by our best-fit $\beta$ at the median mass in each study $\tau_{s,\textrm{SF}}(M_*,t)$, taken from the best fit including all galaxies listed in \citet{Speagle2014}.

\begin{figure}
\vspace{-90pt}
\plotone{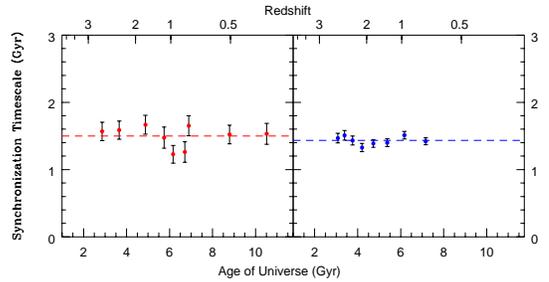}
\caption{Synchronization timescales $\tau_s$ for (red, left) star formation and (blue, right) quasar accretion as a function of time with best-fit linear time dependence.  Star formation timescales are calculated using previously published studies at different redshifts, and quasar timescales are calculated using the Sloan Digital Sky Survey quasar catalog.  $\tau_s$ is consistent with being time-independent in both panels, which is inconsistent with these processes being driven by stochastic events such as major mergers.}
\vspace{+10pt}
\label{fig:syn_both}
\end{figure}
Finally, each study includes star-forming galaxies over a range of redshifts in order to produce a sufficiently large sample for analysis.  Thus, the observed $\sigma$ is produced by a convolution of (1) the scatter in SFR at fixed redshift, (2) the underlying distribution average SFR over that redshift range, (3) associated measurement errors, and (4) cross-correlational scatter between varying SFR indicators themselves. As (3) is generally small, \citet{Speagle2014} deconvolves (2) and (4) to produce a ``true'' $\sigma$ (1), which is used to calculate the value of $\tau_{s,\textrm{SF}}(t)$ given in Table \ref{table:sfr}.

For each redshift range, $t$ is calculated corresponding to the center of the redshift range, and $\dot{\overline{\Psi}}(M_*,t)$ is calculated using the center of the observed mass distribution, since median stellar masses are not reported.  The resulting 9 measurements of $\tau_{s,\textrm{SF}}(t)$ (Fig. \ref{fig:syn_both}) are best-fit by 
\begin{equation}
\frac{\tau_{s,\textrm{SF}}(t)}{1\textrm{ Gyr}} = (1.58 \pm 0.16) - (0.010 \pm 0.025) \frac{t}{1\textrm{ Gyr}},
\end{equation}
which is consistent with a time-independent $\tau_{s,\textrm{SF}} \sim 1.51$ Gyr, estimated at the center of the observed range. If we include all available measurements minus those excluded via criteria 2 and 7 (see Table \ref{table:sfr}), the best fit is instead $\frac{\tau_{s,\textrm{SF}}(t)}{1\textrm{ Gyr}} = (1.19 \pm 0.20) - (0.019 \pm 0.033) \frac{t}{1\textrm{ Gyr}}$, consistent with our best fit. We find both of these are consistent with being fit by a function with constant slope, which gives us best fits of $\tau_{s,\textrm{SF}} = (1.50 \pm 0.18)$ Gyr and $\tau_{s,\textrm{SF}} = (1.31 \pm 0.71)$ Gyr for our best (9 observations) and full (24 observations) sample, respectively.

We note that this time-independence is a strong requirement for a ``main sequence''-like model of galaxy evolution, but is inconsistent with a strong, stochastically-dominated model for star formation.

\section{Quasar Synchronization Timescales}
\label{sec:qso}

We calculate synchronization timescales for quasars using the Sloan Digital Sky Survey (SDSS) DR8 quasar catalog \citep{DR8}.  For quasars in SDSS the galactic mass cannot be measured directly, but virial mass estimators for the central black hole \citep{Vestergaard2006,McLure2002,McLure2004} have allowed \citet{Shen2011} to produce a value-added catalog with black hole masses.  Just as star-forming galaxies have a tight relationship between stellar mass and an indicator of stellar mass growth (star formation), quasars have also been shown to have a tight relationship between black hole mass ($M_\textrm{BH}$) and an indicator of accretion rate (bolometric luminosity; $L$) \citep{Steinhardt2010a,Steinhardt2010b}.  Therefore, we calculate synchronization timescales for quasar accretion using quasars with a common black hole mass to define an ensemble of galaxies, then comparing the observed bolometric luminosity of individual quasars to the average bolometric luminosity for that black hole mass as a function of cosmic epoch.

As with star formation, it is necessary as an intermediate step to find the evolution of average quasar luminosity $\overline{L}(M_\textrm{BH},t)$ at fixed black hole mass, even though the mass of individual supermassive black holes is increasing with time.  Then, if the standard deviation for quasar luminosity is $\sigma(M_\textrm{BH},t)$,
\begin{equation}
\tau_{s,\textrm{QSO}}(M_\textrm{BH},t) = \frac{\sigma(M_\textrm{BH},t)}{\dot{\overline{L}}(M_\textrm{BH},t)}.
\label{eq:qsotau}
\end{equation}

First, we calculate $\overline{L}(M_\textrm{BH},t)$ by finding the average luminosity for quasars at fixed mass at each redshift where the lowest luminosity quasars lie above the SDSS detection threshold.  At fixed mass, $L(M_\textrm{BH},t)$ is well described a log-linear $\log L(M_\textrm{BH},t) = \alpha + \beta t$ (Table \ref{table:qsol}).

\begin{deluxetable}{l c c}
\tablewidth{0pt}
\setlength{\tabcolsep}{2pt}
\tabletypesize{\footnotesize}
\tablecaption{Best-fit average quasar luminosity $\log L(M_\textrm{BH},t) = \alpha + \beta t$ for different mass ranges using the \citet{Shen2011} quasar catalog \label{table:qsol}}
\tablehead{
\colhead{$\log M_{\textrm{BH}}$ (solar)} & 
\colhead{$\alpha$ (dex)} & 
\colhead{$\beta$ (dex / Gyr)}
} 
\startdata
8.5 -- 8.7 & $44.84 \pm 0.02$ & $0.146 \pm 0.005$ \\
8.7 -- 8.9 & $44.88 \pm 0.01$ & $0.153 \pm 0.004$ \\
8.9 -- 9.1 & $44.86 \pm 0.01$ & $0.165 \pm 0.004$ \\
9.1 -- 9.3 & $44.09 \pm 0.02$ & $0.150 \pm 0.006$ \\
9.3 -- 9.5 & $44.96 \pm 0.03$ & $0.170 \pm 0.009$
\enddata
\end{deluxetable}

As for star formation, the measured standard deviation between quasars at fixed $M_\textrm{BH}$ and cosmic epoch is calculated by combining objects over a narrow range of redshift, and $\overline{L}(M_\textrm{BH},t)$ evolves over that range.  Therefore, the measured $\sigma$ is a convolution of the spread induced by changing $\overline{L}$, errors in measuring $M_{BH}$ and $L$, and the residual ``true'' $\sigma$ that descibes the true standard deviation of quasar luminosities at fixed $M_{BH}$ and cosmic epoch.

The evolution of $\overline{L}$ over a narrow redshift range is determined using the best fits shown in Table \ref{table:qsol}.  Although measurement errors provide a negligible to $\sigma$ for star-forming galaxies, they provide a significant contribution to $\sigma$ for quasar accretion.  Virial masses for supermassive black holes have a statistical uncertainty that may be as high as $0.4$ dex \citep{Vestergaard2006}, but appears to be closer to $0.21$ dex for H$\beta$-based masses and $0.15$ dex for Mg{\small II}-based masses \citep{Steinhardt2010c}.  Thus, some of the measured $\sigma$ for quasars is also induced by incorrectly-measured $M_{\textrm{BH}}$, since at fixed redshift, $\overline{L}(M_{\textrm{BH}}) \propto M_{\textrm{BH}}^{0.5 - 0.8}$ \citep{Steinhardt2010b}.  Correcting for both these effects produces a final ``true'' $\sigma$ and, using Equation \ref{eq:qsotau}, corresponding $\tau_{s,\textrm{QSO}}(M_\textrm{BH},t)$ (Table \ref{table:qsosyn}).

\begin{deluxetable*}{l c c c c c}
\tablewidth{0pt}
\setlength{\tabcolsep}{2pt}
\tabletypesize{\footnotesize}
\tablecaption{Best-fit synchronization timescales $\tau_{\textrm{QSO}}(t)$: $\tau_o$ (measured $\sigma$), $\tau_d$ (redshift range corrected), and $\tau_t$ (range and error corrected) for quasar accretion. \label{table:qsosyn}}
\tablehead{
\colhead{Age of the Universe (Gyr)} & 
\colhead{Redshift range} &
\colhead{$\tau_{s,o}$ (Gyr)} & 
\colhead{$\tau_{s,d}$ (Gyr)} & 
\colhead{$\tau_{s,t}$ (Gyr)} & 
\colhead{$\tau_{s,t}$ err (Gyr)}
} 
\startdata
7.17 & 0.6 -- 0.8 & 1.98 & 1.64 & 1.42 & 0.030 \\
6.17 & 0.8 -- 1.0 & 1.93 & 1.72 & 1.51 & 0.030 \\
5.38 & 1.0 -- 1.2 & 1.76 & 1.62 & 1.40 & 0.027 \\
4.74 & 1.2 -- 1.4 & 1.71 & 1.61 & 1.39 & 0.026 \\
4.20 & 1.4 -- 1.6 & 1.60 & 1.53 & 1.33 & 0.026 \\
3.76 & 1.6 -- 1.8 & 1.67 & 1.62 & 1.43 & 0.028 \\
3.39 & 1.8 -- 2.0 & 1.72 & 1.69 & 1.51 & 0.030 \\
3.08 & 2.0 -- 2.2 & 1.67 & 1.64 & 1.47 & 0.032
\enddata
\end{deluxetable*} 

$\tau_{s,\textrm{QSO}}$ is consistent with being independent of mass at fixed redshift, and therefore we average across mass bins in order to produce a more robust measurement of $\tau_{s,\textrm{QSO}}(t)$.  The best-fit linear $\tau_{s,\textrm{QSO}}(t)$ (Fig. \ref{fig:syn_both}) is 
\begin{equation}
\frac{\tau_{s,\textrm{QSO}}(t)}{1\textrm{ Gyr}} = (1.40 \pm 0.04) + (0.004 \pm 0.011) \frac{t}{1\textrm{ Gyr}}.
\end{equation}
This is consistent with time-independence, meaning that quasar behavior can be described with a constant $\tau_{s,\textrm{QSO}} = 1.43$ Gyr, again estimated at the center of the observed range.  As with star formation, this is a strong requirement for a deterministic model for quasar accretion, but is inconsistent with a stochastically-dominated model. Our best fit value with constant slope is $\frac{\tau_{s,\textrm{QSO}}}{1\textrm{ Gyr}} = (1.43 \pm 0.01)$.


\section{Relationship Between Star Formation and Quasar Accretion}
\label{sec:feedback}

In \S~\ref{sec:sf} and \S~\ref{sec:qso}, we show that both star formation and quasar accretion appear to be deterministic rather than stochastic, with $\tau_{s,\textrm{SF}} \approx \tau_{s,\textrm{QSO}} \equiv \tau_s$.  Further, the redshift evolution of both star-forming galaxies and quasars is dominated by two common behaviors: (1) massive galaxies tend to have been assembled earlier than less massive galaxies and (2) a decline in growth rates (star formation rates and black hole accretion rates) towards lower redshift at fixed mass (cf. \citet{Barger2005,Hasinger2005,Steinhardt2010b}).  It is therefore natural to ask whether these processes might be tightly linked, perhaps due to strong feedback.

\citet{Speagle2014} combine 64 different studies on the star-forming main sequence to produce best-fit relationships for the evolution of star formation rates as a function of redshift for a grid of fixed masses in the approximate range $\log (M_*/M_\odot) = [9.5,11.0]$.  This evolution is combined with that of quasars using the \citet{Shen2011} masses and assuming a 8\% radiative efficiency ($\epsilon = 0.08$) in order to express the quasar luminosity in terms of black hole mass growth.

\begin{figure}
\plotone{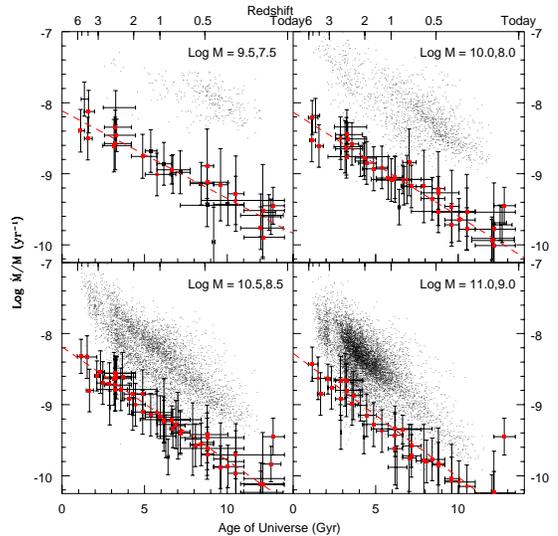}
\caption{The evolution of quasars (objects indicated as black dots) and star-forming galaxies (different studies indicated in red) as a function of redshift in four different mass ranges ($10^9 < M_*/M_\odot < 10^{11}$ and $10^7 < M_{BH}/M_\odot < 10^9$).   Both quasar accretion rates and star formation rates (defined as $\dot{M}$ here) decline over time at fixed mass, and do so in a similar manner.  If the quasar duty cycle is less than 100\%, supermassive black hole growth rates would be correspondingly slower.  In each panel, the mass ratio between the corresponding galaxies and supermassive black holes is 2.0 dex.  Choosing a different mass ratio would result in a similar figure.}
\label{fig:qsosfr}
\end{figure}

We can compare the redshift evolution of quasars and star-forming galaxies by considering the decline in growth rates with redshift (Fig. \ref{fig:qsosfr}).
Because individual objects grow in mass, the same object will appear in multiple panels.  The slope of that decline therefore is determined by comparing, e.g., star-forming galaxies active at some redshift with a different ensemble of star-forming galaxies active 5 Gyr later (which should have lower final masses due to downsizing).  

If galaxies follow a deterministic track such that both (1) there is a constant time gap between star formation and quasar activity and (2) mass growth in both cases scales the same way with ``turnoff'' mass (i.e. the mass at which objects ``turn off'' these relations due to, e.g., quenching), as it would if the two are closely linked to either each other of the remaining mass of unused gas, these slopes will be approximately the same, as is observed in Fig. \ref{fig:qsosfr}.  Otherwise, these similar slopes would be highly concidental.  Thus, there indeed appears to be a tight relationship between star formation and supermassive black hole accretion for a broad range of galaxies and redshifts.  

We note that choosing a different mass ratio between panels in Fig. \ref{fig:qsosfr} would continue to yield similar slopes, but with a different offset between quasars and star-forming galaxies in any given panel.  The offset appears to imply that quasars spend less time accreting mass than their hosts spend forming stars, and might indicate a difference in duty cycles.  Factoring in duty cycles, quasars and star-forming galaxies might then occupy an identical locus if the right mass ratio is chosen.  Alternatively, linked asynchronous evolution might indeed involve different amounts of time spend in quasar and star-forming galaxy states.

In order to quantify this evolution, we fit the quasar data using the same methodology as \citet{Speagle2014}, fitting quasar $\overline{L}(t)$ relations in bins of fixed $M_\textrm{BH}$ of 0.1 dex and the resulting changes in the parameters as a function of mass to derive linear time-dependent coefficients governing how the black hole accretion rate (BHAR) relates to the black hole virial mass. Our best fits of the form $\log M = \alpha(t) \log M + \beta(t)$, using only mass bins for which the majority of virial masses have been derived using Mg{\small II}, are:
\begin{eqnarray}
\log \textrm{BHAR} & = & \left[(0.52 \pm 0.02) - (0.030 \pm 0.002)\,t \right] \times \log M_\textrm{BH} \nonumber \\
& & - \left[(3.53 \pm 0.16) - (0.10 \pm 0.03)\,t \right]. 
\end{eqnarray} For comparison, our best fits to the evolution of star-forming galaxies from \citet{Speagle2014} is:
\begin{eqnarray}
\log \textrm{SFR} & = & \left[(0.84 \pm 0.02) - (0.026 \pm 0.003)\,t \right] \times \log M_* \nonumber \\ 
& & - \left[(6.51 \pm 0.24) - (0.11 \pm 0.03)\,t \right].
\end{eqnarray}
The time evolution of these relations, $\dot{\alpha}_{\textrm{BH}} = - 0.030 \pm 0.002$ and $\dot{\beta}_{\textrm{BH}} = + 0.10 \pm 0.03$ for quasars and $\dot{\alpha}_{\textrm{SF}} = - 0.026 \pm 0.003$ and $\dot{\beta}_{\textrm{SF}} = + 0.11 \pm 0.03$ for SFGs are consistent with being identical. This implies that, although the slopes and normalizations -- and hence the actual fractional growth rates as well as the precise relationship between mass and fractional growth -- for star-forming galaxies and quasars differ, they still appear to evolve ``in sync'' for the majority of the age of the Universe.

The offset between galaxies and quasars in Fig. \ref{fig:qsosfr} is determined by how individual galaxies grow in mass between panels.  It would be constant if both were growing at the same fractional rate, since the panels have a constant mass ratio.  The observed evolution in offset (but not in comparative slopes) is due to the higher fractional mass growth in quasars.

\subsection{Mass Evolution and Turnoff}
\label{sec:turnoff}

The relationship between quasars and star formation shown in Fig. \ref{fig:qsosfr} could have been produced with any of a family of scaling relationships, depending upon the mass ratio between quasars and corresponding star-forming galaxies.  As shown below, $z \sim 0$ galaxies are unable to constrain this ratio.  However, a strong constraint comes from examining the details of mass evolution and turnoff.

At fixed redshift, the number density of (purely) star-forming galaxies as a function of stellar mass is nearly constant for a wide mass range, but declines sharply on the high-mass end as the proportion of quiescent galaxies sharply rises (cf. \citet{Brinchmann2004}, \citet{Salim2007}, \citet{Whitaker2012}, \citet{Moustakas2013}). For the purposes of this paper, this ``turnoff'' mass $M_T$ is defined in several ways, based on the way that SFMS has been represented in the literature. This is divided into a low and high estimate, depending on how this turnoff is defined: closer to where the distribution of galaxies bifuricates and/or levels off in the stellar mass-SFR plane (low), or where the total number of star-forming galaxies sharply declines (high). 

For studies which include mass functions or those where we can access the data directly \citep{bundy+06,Ilbert2013,Steinhardt2014}, $M_T$ is defined as the mass at which the number density has declined to $\sim$25\% (low) or $\sim$10\% (high) of its peak value in that redshift range. For studies that report galaxy distributions as contours in the $M_*$-SFR plane \citep{Brinchmann2004,zamojski+07,Salim2007,Moustakas2013}, we take $M_T$ as the approximate midpoint of the 50\% contours at which the previously unimodal distribution bifuricates or where star-forming galaxies become composite star-forming/AGN sources (low), or the right boundary of the 50\% contours (high). For data where only individual star-forming galaxies (or their medians) are reported \citep{Noeske2007}, we determine the turnoff mass as the position where there is an observed break and/or ``flattening'' in slope from one of approximately unity in the $M_*$-SFR plane to one closer to zero (low) or where the number density of star-forming galaxies is observed to sharply decline (high). We do the same for stacked data \citep{chen+09,oliver+10} using statistics on the number (and type) of objects in each bin. For quasars \citep{Shen2011}, we simply define the turnoff mass where the number density of quasars in the mass-luminosity plane is observed to sharply decline -- this definition is kept the same in both cases to illustrate possible variation in the turnoff relations according to the chosen parametrization. The derived turnoff masses are listed in Table \ref{tab:turnoff}.

\begin{deluxetable*}{l c c c c c c c}
\tablewidth{0pt}
\setlength{\tabcolsep}{2pt}
\tabletypesize{\footnotesize}
\tablecaption{Turnoff Masses $M_T$ derived from the literature \label{tab:turnoff}}
\tablehead{
\colhead{Paper} & 
\colhead{$z_\textrm{low}$} &
\colhead{$z_\textrm{mid}$} & 
\colhead{$z_\textrm{high}$} & 
\colhead{$t_\textrm{mid}$} & 
\colhead{$\log M_T$ ($M_\odot$) (high)} & 
\colhead{$\log M_T$ ($M_\odot$) (low)} &
\colhead{Def.}
} 
\startdata
\textbf{Star formation data} \\
Noeske+07 & 0.2 & 0.325 & 0.45 & 9.817 & 10.9 & 10.4 & 3,4 \\
-- & 0.45 & 0.575 & 0.70 & 7.912 & 11.2 & 10.6 & 3,4 \\
-- & 0.70 & 0.775 & 0.85 & 6.760 & 11.3 & 10.8 & 3,4 \\
-- & 0.85 & 0.975 & 1.10 & 5.848 & 11.4 & 10.9 & 3,4 \\
Steinhardt+14\nocite{Steinhardt2014} & 4.0 & 5.0 & 6.0 & 1.152 & 11.6 & 11.3 & 1 \\
Chen+09\nocite{chen+09} & 0.005 & 0.11 & 0.22 & 12.040 & 10.75 & 10.25 & 4,5 \\
Oliver+10\nocite{oliver+10} & 0.0 & 0.1 & 0.2 & 12.161 & 10.625 & 10.125 & 4,5 \\
Santini+09 & 0.3 & 0.45 & 0.6 & 8.789 & 10.8 & 10.3 & 3,4 \\
-- & 0.6 & 0.8 & 1.0 & 6.635 & 11.1 & 10.4 & 3,4 \\
-- & 1.0 & 1.25 & 1.5 & 4.875 & 11.3 & 10.8 & 3,4 \\
-- & 1.5 & 2.0 & 2.5 & 3.223 & 11.4 & 11.0 & 3,4 \\
Zamojski+07\nocite{zamojski+07} & 0.55 & 0.675 & 0.8 & 7.302 & 10.8 & 10.5 & 2 \\
Salim+07 & 0.005 & 0.11 & 0.22 & 12.040 & 10.9 & 10.2 & 2 \\
Bundy+06\nocite{bundy+06} & 0.4 & 0.55 & 0.7 & 8.077 & 10.7 & 10.45 & 1 \\
-- & 0.75 & 0.875 & 1.0 & 6.278 & 11.0 & 10.7 & 1 \\
-- & 1.0 & 1.2 & 1.4 & 5.032 & 11.2 & 10.95 & 1 \\
Moustakas+13 & 0.2 & 0.25 & 0.3 & 10.520 & 10.7 & 10.45 & 2 \\
-- & 0.3 & 0.35 & 0.4 & 9.598 & 10.8 & 10.4 & 2 \\
-- & 0.4 & 0.45 & 0.5 & 8.789 & 10.8 & 10.4 & 2 \\
-- & 0.5 & 0.575 & 0.65 & 7.912 & 10.9 & 10.6 & 2 \\
-- & 0.65 & 0.725 & 0.8 & 7.023 & 11.0 & 10.7 & 2 \\
-- & 0.8 & 0.9 & 1.0 & 6.166 & 11.2 & 10.9 & 2 \\
Brinchmann+04\nocite{Brinchmann2004} & 0.0 & 0.1 & 0.2 & 12.161 & 10.5 & 10.0 & 2 \\
Ilbert+13\nocite{Ilbert2013} & 0.2 & 0.35 & 0.5 & 9.598 & 10.8 & 10.5 & 1 \\
-- & 0.5 & 0.65 & 0.8 & 7.447 & 10.9 & 10.5 & 1 \\
-- & 0.8 & 0.95 & 1.1 & 5.951 & 11.0 & 10.6 & 1 \\
-- & 1.1 & 1.3 & 1.5 & 4.726 & 11.0 & 10.8 & 1 \\
-- & 1.5 & 1.75 & 2.0 & 3.658 & 11.2 & 11.0 & 1 \\
-- & 2.0 & 2.25 & 2.5 & 2.866 & 11.3 & 11.2 & 1 \\
-- & 2.5 & 2.75 & 3.0 & 2.321 & 11.3 & 11.1 & 1 \\
-- & 3.0 & 3.5 & 4.0 & 1.770 & 11.4 & 11.1 & 1 \\
\textbf{Quasar data} \\
This paper & 0.2 & 0.3 & 0.4 & 10.044 & 9.3 & -- & 4 \\
-- & 0.4 & 0.5 & 0.6 & 8.421 & 9.4 & -- & 4 \\
-- & 0.6 & 0.7 & 0.8 & 7.160 & 9.5 & -- & 4 \\
-- & 0.8 & 0.9 & 1.0 & 6.166 & 9.6 & -- & 4 \\
-- & 1.0 & 1.1 & 1.2 & 5.371 & 9.7 & -- & 4 \\
-- & 1.2 & 1.3 & 1.4 & 4.726 & 9.8 & -- & 4 \\
-- & 1.4 & 1.5 & 1.6 & 4.197 & 9.8 & -- & 4 \\
-- & 1.6 & 1.7 & 1.8 & 3.756 & 9.9 & -- & 4 \\
-- & 1.8 & 1.9 & 2.0 & 3.386 & 10.0 & -- & 4
\enddata
\tablecomments{Col. 1: Papers from which $M_T$ are drawn. Col. 2-4: Lower bound, midpoint, and upper bound of redshift range reported. Col. 5: Age of the Universe in Gyr at the center redshift reported. Col. 6-7: Derived turnoff masses $M_T$. These are identical for quasars, but differ for galaxies. Col. 7: The definition used for $M_T$, which are as follows: 1 -- the mass at which the number density has declined to $\sim$25\% (low)/$\sim$10\% (high) of its peak value in within the reported redshift range, 2 -- the midpoint of the 50\% contours at which the previously unimodal distribution bifuricates or where star-forming galaxies become composite star-forming/AGN sources (see, e.g., \citet{Salim2007}) (low)/the right boundary of the 50\% contours (high) contours, 3 -- the position where there is an observed break and/or ``flattening'' in slope from one closer to unity in the $M_*$-SFR plane to one closer to zero (used for low $M_T$'s), 4 -- where the number density of star-forming galaxies is observed to sharply decline (used for quasars and high $M_T$'s), 5 -- the position where there is an observed break in slope in stacked data.}
\end{deluxetable*} 

Although these definitions do vary, we note that they are intended to be different methods of parametrizing the same general behavior (the quenching/cessation of star formation/black hole accretion) -- while the quantitative results might be subject to unknown systematic errors, the general trends these $M_T$ values show as a function of time should remain robust. Note that, in all cases, we only consider the high-mass turnoff because at high masses, star-forming galaxies and quasars lie well above detection thresholds, while observed low-mass turnoff may be an artifact of survey selection.  This means we are able to measure the synchronization of SF and quasar turnoff, but not of ``turnon''.

\begin{figure*}
\plotone{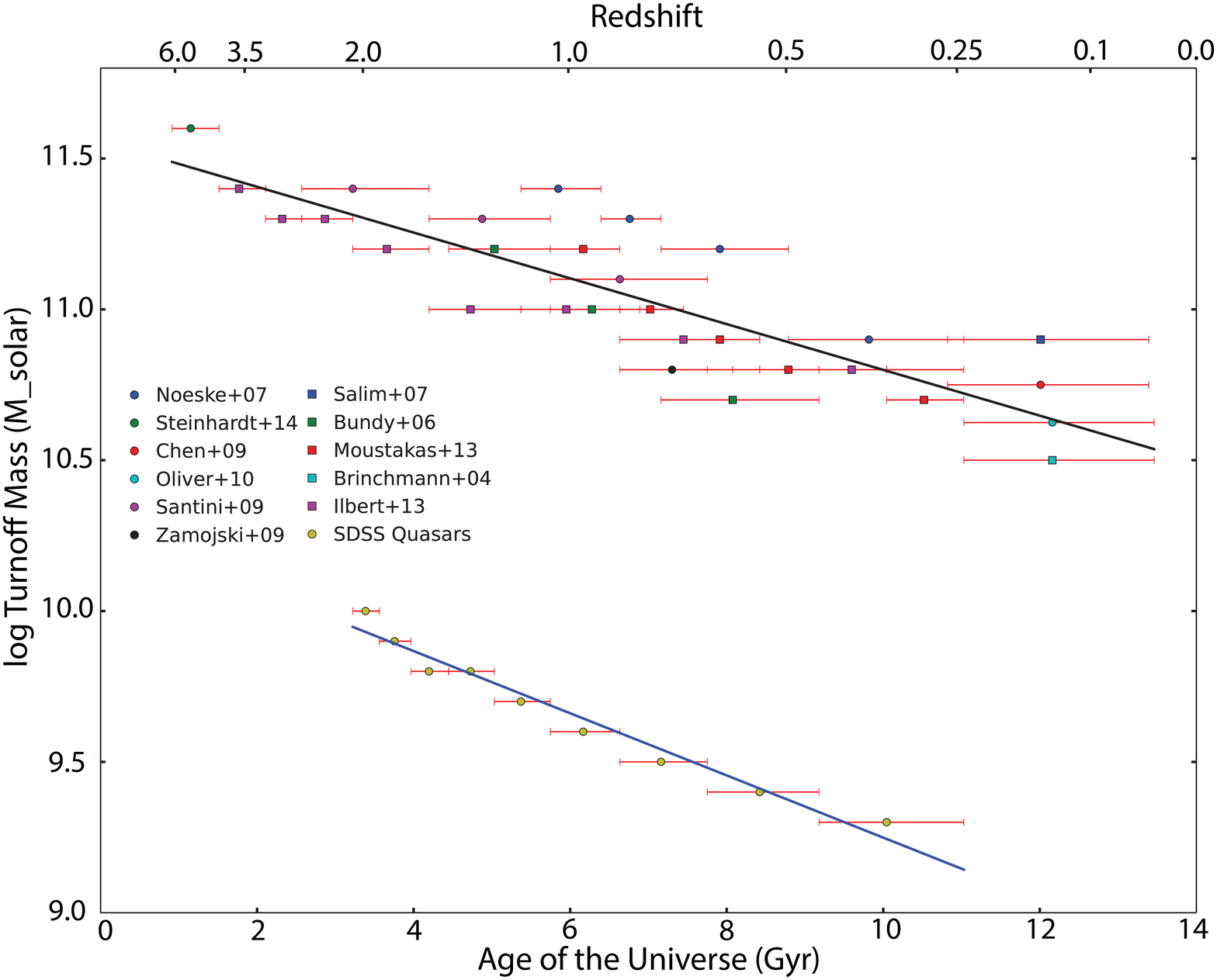}
\caption{Comparison of turnoff masses for star formation (red) and quasar accretion (blue) as a function of cosmic time.  Turnoff masses for star formation come from a variety of studies in the existing literature, and turnoff masses for quasars are calculated using the Sloan Digital Sky Survey quasar catalog.  Both processes show mass downsizing, with higher-mass galaxies being assembled at earlier times.  The best-fit slopes are consistent with being identical, implying a strong connection between the two processes.  However, at fixed redshift, the mass ratio $M_*/M_\textrm{BH}$ is well below the 500:1 ratio reported at $z \sim 0$.}
\label{fig:turnoff}
\end{figure*}

Consistent with previous reports of downsizing, the turnoff mass for both star-forming galaxies and quasars decreases towards later times (Fig. \ref{fig:turnoff}).
In both cases, the turnoff is well-fit by linear time evolution:
\begin{eqnarray*}
\log M_{T,\textrm{SF}}/M_\odot & = (11.56 \pm 0.06) - & (0.08 \pm 0.02) t \\
\log M_{T,\textrm{QSO}}/M_\odot & = (10.28 \pm 0.06) - & (0.10 \pm 0.02) t,
\end{eqnarray*}
where the SF turnoff relation is taken from the ``high'' $M_T$ values. If we had instead chosen the ``low'' values, it would instead be $\log (M_{T,\textrm{SF}}/M_\odot) = (11.35 \pm 0.05) - (0.10 \pm 0.02) t$.

The slopes for both turnoff masses are consistent with being identical, so that at every redshift, the ratio between the stellar mass of turnoff SF galaxies to the black hole mass of turnoff quasars is approximately constant.  Near redshift 0, the same is true of quiescent galaxies \citep{Magorrian1998,Haering2004, McConnell2013}.  However, at fixed redshift, the observed ratio of turnoff masses between galaxies and their central black holes ranges from 20:1 to 30:1 in the above fits, while for quiescent galaxies it is approximately 500:1.  Therefore, the SF galaxies and quasars turning off at the same redshift are mismatched and likely are different ensembles of galaxies with different halo masses.  Matching the observed Magorrian relation ratio of $M_*/M_{BH}$ requires a larger $M_*$ than is observed to turn off concurrently with quasars.  This increased $M_*$ might be produced by a combination of star formation turning off at higher redshift than quasars (turnoff $M_*$ increases towards higher redshift) and ``hidden'' star formation, in which $M_*$ increases but the galaxy is never selected as star-forming.  Therefore, the correct mass ratio between corresponding star-forming galaxies and quasars is likely larger than 30:1, with the quasar phase occurring at later times, but lower than 500:1.

We note that ``hidden'' star formation must be a significant component, since the largest quasars have $M_{BH} > \sim 10^{9.8} M_\odot$ and therefore would lie in galaxies of $M_* \sim 10^{12.5} M_\odot$, while the most massive star-forming galaxies observed at any redshift are turning off with $M_* < \sim 10^{11.5} M_\odot$.  Thus, the mass ratio is probably no larger than 1.7 dex, or 50:1. This is in good agreement with the results from \citet{Matsuoka2014}, who find that stellar masses of quasar host galaxies at $z < 0.6$ are lower than predicted from the Magorrian relation by $\sim 0.8$\,dex (see their Fig.~16).


\section{Discussion}
\label{sec:discussion}

Observations of star-forming galaxies and quasars at a wide range of redshifts indicate a tight relationship between galaxies of a common mass and redshift.  In this paper, we introduced a quantitative measure of the extent to which different galaxies of a common mass are evolving synchronously.  Using 9 different star-forming galaxy observations from the literature and SDSS quasar observations spanning $0 < z \lesssim 6$, this synchronization timescale appears to be a constant $\tau_s \sim 1.5$ Gyr.  In other words, choosing the ensemble of all galaxies at any fixed mass and any stage of star formation of quasar accretion which they all go through, the variance in times at which individual galaxies of that mass go through that stage is approximately 1.5 Gyr.  

Because of the prevalence of stochastic processes in galactic evolution, it might have been expected that galaxies of a common mass are more similar to each other at high redshift than at low redshift.  This is inconsistent with our measurements of $\tau_s$, which instead indicate that galaxies are equally well synchronized at high and low redshift, implying that an ensemble of similar high-redshift galaxies will have similar histories.

What, then, is the role of mergers and other environmental factors if galaxies are following a common track?  One possibility is that this common history is an attractor solution, with mergers temporarily increasing the gas supply and feedback relaxing the galaxy back to the track (cf. \citet{Peng2014}).  If so, the 1.5 Gyr timescale would be determined by gas dynamics and the details of that feedback mechanism.  For most galaxies in the range we observe, 1.5 Gyr is $\sim$5 dynamical times.  Another possibility would be such a relationship resulting from averaging a large number of smaller mergers \citep{Fakhouri2008,Munoz2014}.  In this model, individual galaxies might sometimes have more and sometimes less star formation than average at different points in their history.

Alternatively, note that for an ensemble of halos of a fixed mass, there will be a range of virialization times, since they might form from larger, less overdense regions or smaller, more overdense ones.  The synchronization time for forming these halos varies only slightly with mass, and is approximately 1.5 Gyr \citep{Press1974,Haiman1997,Sasaki1994}.  Thus, we might also consider the opposite extreme: developing a model for galaxies dominated by very strong feedback, such that stochastic processes play a negligible role, and galaxies follow a deterministic track.  In such a model, $\tau_s$ is laid down by the spread initial virialization times, so that galaxies with more star formation than average for their redshift are galaxies that formed later, and will continue to lie above the star-forming main sequence for their entire history.  

Distinguishing between these two explanations is beyond the scope of this paper.  However, we note that for both types of models, galaxies will have a common history, and an obvious next step is to gain a better understanding of the sequence of phases galaxies go through in this history.  Our methodology sheds light on the relationship between star formation and quasar accretion during that history.  As discussed in \S~\ref{sec:turnoff}, the masses of galaxies and quasars turning off at the same redshift do not lie on the Magorrian $M_\textrm{BH}-M_*$ relation, implying that the observed star formation in galaxies may occur prior to quasar accretion, and that much of a galaxy's stellar mass growth takes place later and is ``hidden'', taking place under conditions that prevent it from being selected as a star-forming galaxy.

\citet{Steinhardt2011} use observed quasar distributions to empirically fit evolutionary tracks for individual supermassive black holes, finding that quasar may live for just 1-2 Gyr, yet have a duty cycle such that they are luminous for all of that time.  Star formation during a quasar phase would indeed be ``hidden'', because a luminous quasar is far brighter than its host galaxy.  Similarly, \citet{Leitner2012} use empirical fits to the star forming main sequence and find that star formation may also be one long, extended phase rather than episodic. \citet{Behroozi2013} find a similar result using empirical fits to all galaxies, especially at higher redshifts.

As a result, we are led to propose a history in which galaxies at fixed mass tightly adhere to a common main sequence, not just for star formation, but including all of the following phases (Fig. \ref{fig:cartoon}):
\begin{figure*}
\includegraphics[scale=0.60,angle=270]{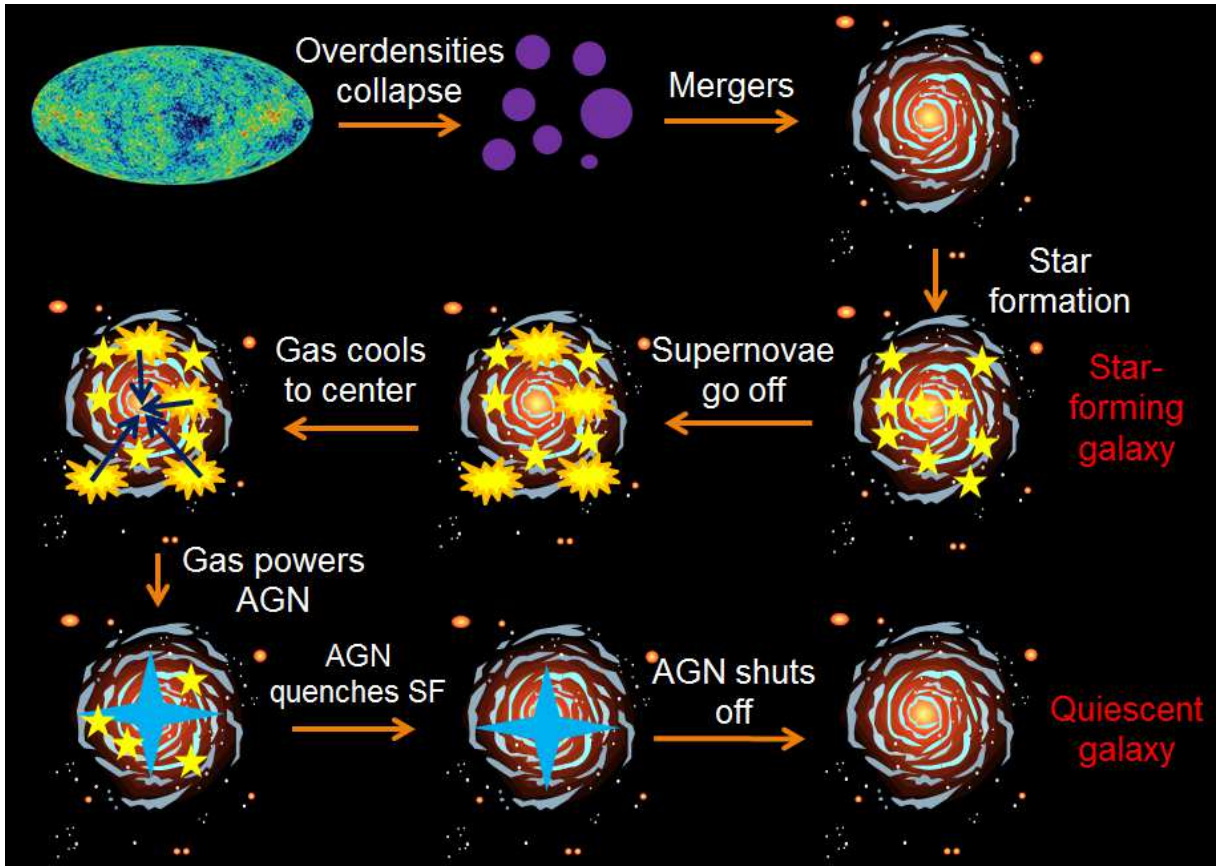}
\caption{The major phases in a deterministic model for galactic evolution.}
\label{fig:cartoon}
\end{figure*}
\begin{enumerate}
\item{{\bf Mergers} in the early universe produce virialized halos.  This produces an initial $\tau_s$ of approximately 1.5 Gyr.}
\item{{\bf Star formation} occurs, not stochastically in large, episodic starbursts but rather as one, long quasi-continuous star-forming phase.}
\item{{\bf Quasars turn on}, powered by that gas.  This is accompanied by ``hidden'' star formation, but because the quasar is always luminous during this phase, these are not detected as star-forming galaxies or included in studies of SFMS.}
\item{{\bf Quenching} of star formation by feedback from the central active galactic nucleus (AGN).  If quasars are active in total for approximately 2 Gyr, then ``hidden'' star formation has increased $M_*$ by a factor of $\sim$10 before quenching in order to lie on the $M_{BH}-M_*$ relation at redshift 0.}
\item{{\bf Quasars turn off} as they run out of fuel.}
\item{{\bf Bulges form}, either at the end of or throughout this process \citep{Abramson2014}, producing the $M-\sigma$ relation \citep{msigma1,msigma2} at low redshift.}
\item{{\bf Quiescence} as observed at $z=0$.}
\end{enumerate}

\vskip.5cm

The entirety of the history given by this model cannot be tested directly for individual galaxies, because data are only available at a single redshift. However, several key predictions seem to hold up. For instance, such a model would predict that, in addition to hidden star formation, quasars should be preferentially hosted in galaxies with large stellar masses. Both of these predictions are consistent with the results seen in \citet{Matsuoka2014} (for direct AGN/host galaxy image decomposition) and \citet{Salim2007} (using the BPT diagram). The results of \citet{Salim2007} further indicate that not only are BHs hosted in more massive star-forming galaxies, but that their activity is strongest at lower masses (in composite AGN/SF systems) and declines at higher masses (in AGN-dominated systems) (see their Fig.~18 and~19). These results both are consistent with stages 3-5. 

In addition, \citet{Schiminovich2007} find that a significant fraction of galaxies with sSFR above those on the SFMS are bulge-dominated, and find that a significant fraction of these galaxies are likely be experiencing a final episode of star formation that can explain the growth rate of quiescent galaxies at $z \sim 0$. This establishes a strong link between eventual quenching and bulge formation, which is consistent with stages 5-7 (see also their Fig.~23 and \citet{Abramson2014}).  Furthermore, \citet{Shim2011} find that at $z \gtrsim 4$ star-forming galaxies show strong evidence for extended star formation timescales, indirect evidence supporting star formation mechanisms that fit our stage 2. Finally, a number of studies \citep{Maraston2010,Lee2011,Papovich2011,Gonzalez2012,Speagle2014} have found strong evidence that galaxies at $z \gtrsim 2$ are well fit by extended, rising star formation histories rather than exponential bursts, again in support of stage 2.

This sketch of this history is clearly overly simplistic, as there are many complex processes involved in galaxy formation, star formation, supermassive black hole accretion, feedback, etc.  In this work, we have outlined the requirements of this model during each stage, but specific physical models for each of these complex processes meeting these requirements are necessary to produce a working model.  Nevertheless, this surprisingly simple picture would be consistent with the observed evolution of galaxies and their central black holes at all observed redshifts, while a strong stochastically-dominated scenario appears not to be.

\subsection{The Role of  Mergers}

Although most star forming galaxies lie on the main sequence, it should be noted that at $z \lesssim 2$, 8-14\% are observed to lie at much higher star formation rates \citep{Rodighiero2011,Elbaz2011,Whitaker2012,Sargent2012}. Such galaxies appear not to fit easily into our common history, as they do not lie on the main sequence which prompted its development.

Follow-up observations indicate that these galaxies are typically in the midst of a major merger \citep{Rodighiero2011,Elbaz2011}, with a mass ratio close to unity.  The key question that cannot be directly answered observationally is what happens following one of these major mergers.  There is strong evidence that ultra-luminous infrared galaxies (ULIRGs) such as these become elliptical \citep{Genzel2001,vanderwel2009}, but their subsequent history is less certain.

We can rule out the possibility that these galaxies continue as high-star formation rate outliers, since we would see them in star formation studies.  Although it is possible that they return to this deterministic main sequence, a more likely answer is that this sort of major merger accelerates the depletion of gas, forming stars and a supermassive black hole rapidly, then becoming quiescent and by low redshift producing the observed population of old, red elliptical galaxies (cf. \citet{Whitaker2012,Ilbert2013,Toft2014}).  If at any given time, 1.5\% of galaxies are pulled off the main sequence permanently due to a merger that takes place, e.g., over 200 Myr \citep{Lotz2008}, then after 10 Gyr, 47\% of galaxies will never have had such a merger and will continue to populate the main sequence.  If the dynamical timescale for a merger is closer to 1 Gyr, 86\% of galaxies stay on this deterministic track.  Thus, major mergers might be important to the story for some individual galaxies, particularly in large clusters, yet insignificant for the remainder. 

As noted by \citet{Shim2011}, however, at higher redshifts mergers might have a negligible impact on galaxy SFRs: their spectroscopic sample contains $\sim 50\%$ visually classified mergers, but has almost no high SFR outliers.  Mergers thus may only play an important role when gas is unavailable -- at high redshift, when gas is readily available, additional influxes of gas will not expedite evolution. However, at lower redshift, where gas is more scarce, a major merger might bring in a large influx of gas (from, e.g., the surrounding halo), replicating conditions at higher redshift (see also Khabiboulline et al., subm.). This, coupled with turbulent shocks and other merging phenomena, would lead to a strong ``boost'' in the sSFR in any particular lower redshift merging system.

If most galaxies instead built their mass through a large number of \textit{minor} mergers, this might instead act as a stablization mechanism.  By the central limit theorem, building galaxy properties by combining a large number small objects in the early universe is likely to produce a very uniform set of initial conditions.  Similarly, if stochastic processes in the galaxy are driven by the inflow of fresh gas through mergers, a large number of minor mergers would produce a consistent and universal driving mechanism, which might then lead to galaxies evolving in a universal manner.

In conclusion, a variety of observations -- of different processes, at a broad range of redshifts, and using different techniques -- imply that we should be searching for a simple, deterministic model of galactic evolution in which the history of most individual galaxies follows a common sequence of events.  The strongest of these constraints appears to be the 30:1 ratio between the most massive star-forming galaxies and central black holes at fixed redshift, which appears to require that the last phase of star formation in massive galaxies is somehow hidden.  Using the new methodology described in this paper, we have proposed one possible sequence consistent with all of these observational constraints.  Ours is by no means the only possible sequence, and it should be considered essential to determine which sequence is correct.

\acknowledgements
The authors would like to thank Steve Bickerton, Kevin Bundy, Peter Capak, Marcella Carolla, Martin Elvis, Brian Feldstein, Peter Goldreich, Emil Khabiboulline, Simon Lilly, Dan Masters, Nick Scoville, John Silverman, and Michael Strauss for their valuable comments.  The authors also wish to thank an anonymous referee for helpful recommendations.  JSS was partially supported by the Harvard Office of Career Service's Weismann International Internship Program and the Harvard College Research Program.

\bibliographystyle{apj}

\end{document}